# 07

# PROPUESTA DE PLATAFORMA
## DE PROCESAMIENTO DE DATOS PARA MARKETING DIRECTO
### PROPOSAL OF DATA PROCESSING PLATFORM FOR DIRECT MARKETING DATA

MSc. Jorge Luis Rivero Pérez[1]
E-mail: jlrivero@dei.uc.pt
Dra. C. Yaimara Peñate Santana[2]
E-mail: yaimara.penates@ug.edu.ec
Ing. Pedro Harenton Martínez López[3]
E-mail: pedro.harenton@tecnoazucar.azcuba.cu
[1] Universidad de Coimbra. Portugal.
[2] Universidad Estatal de Guayaquil. República del Ecuador.
[3] Empresa Tecnoazucar. Cienfuegos. Cuba.

RESUMEN

La minería de datos ha sido ampliamente utilizada para identificar potenciales clientes de un nuevo producto o servicio. En el presente artículo se hace un estudio de trabajos previos referidos a la aplicación de metodologías de minería de datos para proyectos de Software, específicamente para proyectos de marketing directo. Se describen varios conjuntos de datos demográficos e históricos de compras de clientes, disponibles para la evaluación de algoritmos en esta área, algunos de ellos muy novedosos y actuales. Se propone una plataforma para el procesamiento de flujos de datos distribuidos, para los procesos de selección de clientes objetivos y la construcción de modelos predictivos de respuesta requeridos; se facilita así varios de los requerimientos funcionales necesarios para los entornos de desarrollo.

Palabras clave:

Flujos de datos, marketing directo, minería de datos.

ABSTRACT

Data mining has been widely used to identify potential customers for a new product or service. In this article is done a study of previous work relating to the application of data mining methodologies for software projects, specifically for direct marketing projects. Several data sets of demographic and historical customer purchases data available for evaluation of algorithms in this area, some of them very new and current are described. The main contribution of this paper is the proposal of a platform for distributed data stream processing for the processes of targeting customers and building predictive models required response; thus facilitating several of the functional requirements for development environments.

# INTRODUCCIÓN

El marketing tradicional realiza promociones a través de diversos canales como noticias en periódicos, radio, etc., pero esas promociones están dirigidas a todas las personas, interesadas o no en el producto o servicio que se promociona. Por lo general este método conduce a grandes gastos y a un bajo índice de respuesta por parte de los posibles clientes. Es por ello, hoy en día, al existir un mercado muy competitivo, el marketing masivo no resulta seguro, de ahí que los especialistas están focalizando los esfuerzos en el marketing directo. Este método estudia las características, necesidades y además selecciona un grupo de clientes como objetivo para la promoción. El marketing directo usa el modelado predictivo a partir de los datos de los clientes, con el objetivo de seleccionar los más propensos a responder a las promociones (Chakrabarti & Mitchell, 2013; Olson & Chae, 2011; Pinto, Mansfield & Rubin, 2011; Raykov & Calantone, 2014; Setnes & Kaymak, 2001).

El marketing directo es un proceso orientado por conjuntos de datos de comunicación directa con clientes objetivos o prospectos (Kacen, Hess & Chiang, 2013; Schweidel & Knox, 2013;(Thorson & Moore, 2013). Hoy en día este enfoque está siendo adoptado por un creciente número de compañías, especialmente por empresas financieras y bancarias (Moro, Cortez & Rita, 2014). Pero a la vez, los especialistas de marketing se enfrentan a la situación de entornos cambiantes. Los conjuntos de datos actuales constituyen computadoras insertan datos dentro de otras haciendo que los entornos sean dinámicos y condicionados por restricciones como capacidad de almacenamiento limitada, capacidad de procesamiento en tiempo real, etc. Esto implica que se requiera adoptar enfoques de procesamiento de Big Data (Fan & Bifet, 2013); de ahí que para lograr establecer y mantener la relación con los clientes, los especialistas han avizorado la necesidad de cambiar los métodos de selección intuitiva de grupos por enfoques más científicos orientados a procesar grandes volúmenes de datos (Karim & Rahman, 2013); (Setnes & Kaymak, 2001), obteniendo respuestas rápidas que permitan seleccionar los clientes que responderán a una nueva oferta de productos o servicios, bajo un enfoque de flujos de datos de procesamiento distribuido.

Varias son las investigaciones que se refieren a aspectos computacionales y teóricos del marketing directo, pero pocos esfuerzos se han focalizado en aspectos tecnológicos necesarios para aplicar minería de datos al proceso de marketing directo. Situación esta que gana en complejidad cuando existen entornos de flujos de datos distribuidos. Los investigadores tienen que dedicar esfuerzos y tiempo a la implementación de entornos de simulación de flujos de datos, tantos demográficos como históricos, de compras, debido a que no se tiene una plataforma que se encargue de tareas como esa y que a la vez sea escalable, lo que permite la incorporación de nuevas variantes de algoritmos para su respectiva evaluación. Ante tal problemática y por la importancia que reviste incluso como base de futuras investigaciones en el área de la minería de flujos de datos para el marketing directo; en la presente investigación se propone una plataforma de procesamiento basada en tecnologías de Software Libre y con un alto nivel de escalabilidad.

# DESARROLLO

La Inteligencia de Negocios es un término sombrilla que incluye metodologías, bases de datos, aplicaciones, herramientas y arquitecturas con el objetivo de usar los datos como soporte para la toma de decisiones por parte de los administradores de negocios (Turban, Sharda, Aronson, y King, 2008). La minería de datos es una tecnología de Inteligencia de Negocios que usa modelos para extraer conocimiento a partir de datos (Witten & Frank 2005). Por lo general, no se encuentran investigaciones referidas específicamente al desarrollo de las arquitecturas diseñadas a partir de la interoperabilidad de varias herramientas de Software para soportar las metodologías capaces de extraer conocimiento a partir de las bases de datos. Las aplicaciones se centran en la implementación de las etapas de las metodologías, pero no se refieren a aspectos tecnológicos muy necesarios para el despliegue de las implementaciones. Esta situación gana relevancia en los escenarios actuales donde cada día se requiere más del análisis de Big Data (Fan & Bifet, 2013), para lo cual resulta imprescindible una arquitectura de procesamiento soportada en la selección tecnologías y estándares actuales que permiten la implementación de soluciones.

Varias investigaciones (Casabayó, Agell & Sánchez-Hernández, 2015; Castro López, 2014; Govidarajan, Karim & Rahman, 2013; Moro et al., 2014; Olson & Chae, 2011; Thompson, Heley, Oster-Aaland, Stastny & Crawford, 2013) se han referido a la aplicación de la minería de datos para el desarrollo de aplicaciones en el marketing directo. En todos los casos se divide el proceso en diferentes etapas y luego se requiere implementar algoritmos de minería de datos para darle solución a cada una de ellas. La Cross-Industry Standard Process for Data Mining (CRISP-DM) es una metodología popular que define una secuencia de seis pasos los cuales permiten construir e implementar modelos a ser usados en entornos reales, sirviendo así como soporte para la toma de decisiones en los negocios (Chapman et al., 2013). CRISP-DM define un proyecto como un proceso cíclico donde varias iteraciones pueden ser usadas para obtener un resultado final más ajustado a los objetivos del negocio.

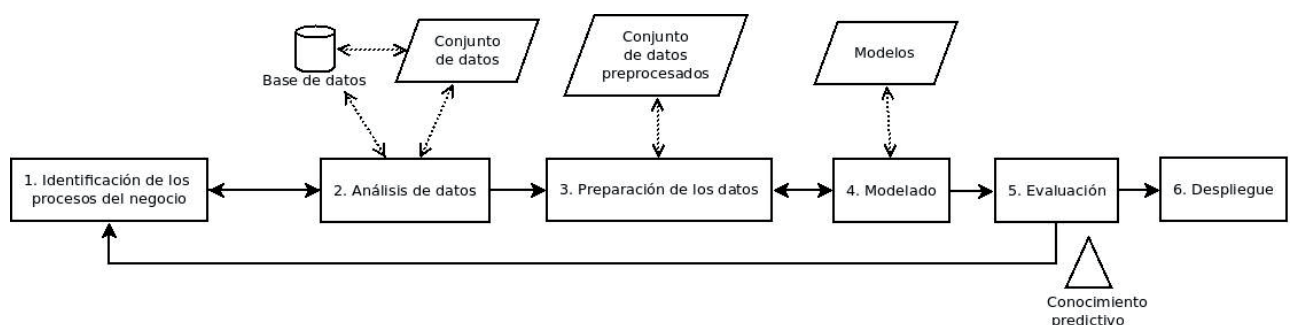

Figura 1. Etapas de la metodología CRISP-DM.

## Metodología

El marketing directo efectivo comienza con una base de datos de clientes, la cual es una colección organizada de datos correspondientes a clientes individuales o prospectos, incluye información geográfica y demográfica (edad, ingresos, miembros de la familia, cumpleaños); psicográfica (actividades, intereses, opiniones) y de comportamiento. Estas bases de datos dan a las compañías y empresas una panorámica de 360 grados de sus clientes atendiendo a su comportamiento. Existen varias compañías encargadas de la gestión de estas bases de datos, entre las que puede citarse *The Cornerstone Group of Companies* (2015), el cual es la compañía más grande de Canadá, encargada de la gestión de los contactos de listas de clientes y de correos electrónicos. Cada año rastrea las tendencias y evalúa la actividad de respuesta en más de 1.100 listas. Grandes volúmenes de esos datos son generados cada día en muchas organizaciones, los cuales pueden ser utilizados para establecer y mantener una relación directa con los clientes para así identificar cuáles serían los grupos de clientes que responderían a ofertas específicas.

Como en cualquier modelado, seleccionar los atributos que serán usados como variables en el modelo de respuesta resulta un paso determinante. Diferentes tipos de bases de datos están disponibles para ser usados por las compañías para hacer marketing directo. La calidad de la base de datos a ser utilizada con propósitos de selección es considerada el aspecto más importante en las campañas de marketing directo.

Una gran variedad de datos, tanto en bases de datos externas como propias de las organizaciones, está disponible para los especialistas de marketing. Las externas contienen información de revisiones independientes de marketing o de estudios demográficos. Existen compañías como *Nielsen Datasets Transforming Academic Research in Marketing* (2015), especializadas en recolectar y mantener datos de ventas para las empresas de marketing.

Una categoría especial son las bases de datos con información demográfica, la cual provee información adicional a nivel de grupo, y son útiles para determinar grupos objetivos que tengan propiedades similares. Contrariamente a las base de datos externas, las internas de las compañías proveen información más relevante y confiable. Respecto al comportamiento de los clientes. Según (Al-Shayea & Al-Shayea, 2014), para propósitos de minería de datos, el histórico de ventas puede ser traducido en atributos basados en medidas de:

- Actualidad (*Recency*).

- Frecuencia (*Frequency*).

- Monetarias (*Money*).

Estos atributos (RFM) son identificados como las tres variables más representativas para el modelado de marketing. Cuando estas variables se usan en combinación se logran resultados predictivos muy precisos y además resultan muy útiles para apoyar la gestión de la relación con los clientes (CRM) (Hollensen, 2015). La ventaja de estas variables es que el comportamiento de los clientes puede ser capturado usando un pequeño número de atributos. Desde el punto de vista del modelado, las variables tienen la ventaja de resumir el comportamiento de compras de los clientes, al usar relativamente una cantidad pequeña de variables. Contrariamente a las variables demográficas, las cuales pueden ser cerca de 100, las RFM tienden a ser 10 o menos para un conjunto de datos. De ahí que la reducción de atributos no resulta un paso crítico cuando se emplean estas variables.

## Conjuntos de datos actuales

Varios son los conjuntos de datos disponibles para la evaluación de los algoritmos en tareas referidas al marketing directo. En este artículo solo se presentan los más actuales en el estado del arte.

Tabla 1. Caracterización de conjuntos de datos para Marketing Directo.

| Conjunto de datos | Características | Instancias | Atributos |
|---|---|---|---|
| *Bank Marketing Data Set* | Multivariado | 45111 | 11 |
| *Insurance Company Benchmark (COIL 1000) Data Set* | Multivariado | 9000 | 86 |

En el *Bank Marketing Data Set* los datos están relacionados con una campaña publicitaria de una institución bancaria portuguesa, dirigida a varios clientes. Existen otros conjuntos de datos como el propuesto por *Nielsen Datasets* (2015) que registran los hábitos de consumo de las personas en los Estados Unidos, los cuales están a disposición de los investigadores académicos. A continuación, en la siguiente tabla son descritos los atributos:

Tabla 2. Descripción de los atributos del Bank Marketing Data Set.

| Número | Atributo | Descripción (Tipo de atributo: posibles valores) |
|---|---|---|
| | | Atributos que describen datos del cliente del banco |
| 1 | Age | Edad del cliente (numérico) |
| 1 | Job | Tipo de empleo (Categórico: 'admin.','blue-collar', 'entrepreneur', 'housemaid', 'management','retired','self-employed','services','student','technician','unemployed','unknown') |
| 3 | Marital | Estado civil (Categórico: 'divorced','married','single','unknown'; note: 'divorced' means divorced or widowed) |
| 4 | education | Nivel educacional (Categórico: 'basic.4y','basic.6y','basic.9y','high.school','illiterate','professional.course','university.degree','unknown') |
| 5 | Default | ¿Tiene su crédito en default? (Categórico: 'no','yes','unknown') |
| 6 | Housing | ¿Tiene casa hipotecada? (Categórico: 'no','yes','unknown') |
| 7 | Loan | ¿Tiene hipotecas? (Categórico: 'no','yes','unknown') |
| | | Atributos relacionados con el último contacto entre la persona y la campaña |
| 8 | Contact | Tipo de contacto para la comunicación (Categórico: 'cellular','telephone') |
| 9 | Month | Último mes que tuvo contacto con la campaña (Categórico: 'jan', 'feb', 'mar', ..., 'nov', 'dec') |
| 10 | day_of_week | Último día de la semana en que tuvo contacto con la campaña (Categórico: 'mon','tue','wed','thu','fri') |
| 11 | Duration | Duración en segundos del último contacto (numeric). |
| | | Otros atributos |
| 11 | campaign | Cantidad de contactos realizados durante la campaña para el cliente (Numérico, incluye el último contacto) |
| 13 | Pdays | Cantidad de días desde el último contacto del cliente con la campaña (Numérico; 999 significa que nunca ha sido contactado) |
| 14 | Previous | Número de contactos que ha tenido el banco con el cliente antes de la nueva campaña (Numérico) |
| 15 | poutcome | Resultado de la campaña de marketing anterior (Categórico: 'failure','nonexistent','success') |
| | | Atributos socio-económicos |
| 16 | emp.var.rate | Variación de la tasa de empleabilidad (Numérico) |
| 17 | cons.price.idx | Índice de precios del consumidor – Indicador mensual (Numérico) |
| 18 | cons.conf.idx | Índice de confiabilidad del consumidor – Indicador mensual (Numérico) |
| 19 | euribor3m | Tasa trimestral de Euribor – Indicador diario (Numérico) |
| 10 | nr.employed | Cantidad de empleos - (Numérico) |
| 11 | Class | El atributo de clase es el *concepto* a ser aprendido. Para este caso sería: ¿Se ha suscrito el cliente a la campaña? (Binario: 'yes','no') |

## Propuesta de arquitectura

### Criterios de selección de tecnologías

Actualmente existen dos enfoques de procesamiento de *Big Data*. Uno de ellos utiliza algoritmos incrementales para su procesamiento y el otro enfoque basa el procesamiento en computación distribuida. Dado que la naturaleza de las fuentes de datos generadas por los hábitos de consumo de los clientes resultan ubicuas, en el artículo se propone un procesamiento de flujos de datos distribuidos, capaz de resolver las tareas de cada una de las etapas de la metodología CRISP-DM. Para ello existen varios *frameworks* y tecnologías que resultan muy útiles, y en algunos casos indispensables para la evaluación y el desarrollo de algoritmos capaces de procesar los datos, tanto para entornos estacionarios como para entornos de flujos de datos. La Figura 1 muestra la clasificación de dichas tecnologías según los entornos de evaluación.

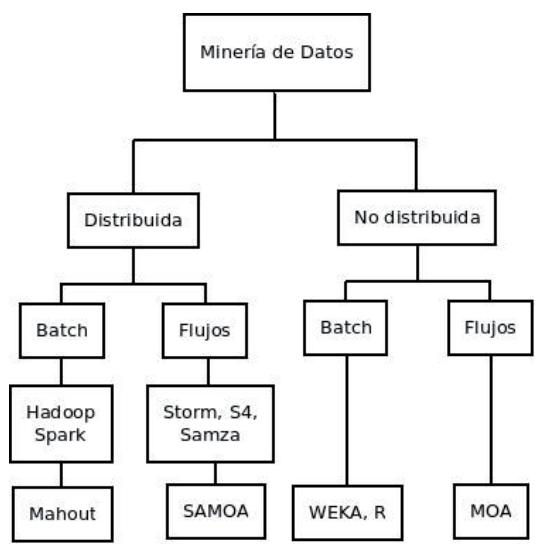

Figura 2. Taxonomía de las herramientas de minería.

### Principios para la selección de tecnologías

Como se refería anteriormente existen varias tecnologías para la minería de datos. Es por ello que para el diseño de la plataforma propuesta se siguieron varios principios de selección:

• Idoneidad: Las herramientas utilizadas en el diseño realizan la función requerida en cada etapa.

• Interoperabilidad: facilidad para integrar y conectar la salida de determinados procesos como entradas de otros.

• Tecnología homogénea: todas las herramientas utilizadas para el diseño son desarrolladas bajo los mismos paradigmas y tecnologías de programación.

• Escalables: pueden ser incorporados nuevos algoritmos para cada una de las etapas de procesamiento de los datos comprendida en la metodología CRISP-DM, así como modificar los ya existentes.

• *Software* Libre: las herramientas seleccionadas cumplen con las cuatro libertades del Software Libre, garantiza así el despliegue de la arquitectura propuesta, como la reproducibilidad de los experimentos.

### Tecnologías seleccionadas

WEKA: Waikato Environment for Knowledge Analysis (Witten & Frank, 2005) soporta varias tareas estándar de minería de datos, especialmente, preprocesamiento de datos, clustering, clasificación, regresión, visualización, y selección. Todas las técnicas de Weka se fundamentan en la asunción de que los datos están disponibles en un fichero plano (flat file) o una relación, en la que cada registro de datos está descrito por un número fijo de atributos (normalmente numéricos o nominales, aunque también se soportan otros tipos). Weka también proporciona acceso a bases de datos vía SQL gracias a la conexión JDBC (Java Database Connectivity) y puede procesar el resultado devuelto por una consulta hecha a la base de datos. No puede realizar minería de datos multi-relacional, pero existen aplicaciones que pueden convertir una colección de tablas relacionadas de una base de datos en una única tabla que ya puede ser procesada con Weka.

MOA: *Massive Online Analysis* (Bifet et al., 2010) es un framework para implementar y evaluar algoritmos de aprendizaje en tiempo real en flujos de datos. MOA incluye una colección de algoritmos para tareas de clasificación y agrupamiento. El framework sigue los principios de green computing empleando los recursos de procesamiento eficientemente, siendo una característica requerida en el modelo de flujos de datos donde los datos arriban a altas velocidades y los algoritmos deben procesarlos bajo restricciones de tiempo y espacio. El framework se integra con WEKA.

SAMOA: *Scalable Advanced Massive Online Analysis* es una plataforma para la minería de flujos de datos distribuidos (Bifet & De Francisci Morales, 2014). Como la mayoría de los frameworks de procesamiento de flujos de datos está escrito en Java. Puede ser utilizado como framework para desarrollo o como biblioteca en otros proyectos de software. Como *framework* permite a los desarrolladores poder abstraerse de tareas referidas a los motores de ejecución, lo que facilita que el código pueda ser utilizado con diferentes motores de procesamiento de flujos, hace que pueda implementarse una arquitectura sobre diferentes motores como Apache Storm, Apache S4 y Samza. SAMOA logra la integración perfecta con MOA y por ende pueden emplearse los algoritmos de clasificación y procesamiento de MOA en entornos de flujos distribuidos. Constituye una plataforma en sí muy útil tanto para investigadores como para soluciones de despliegue en entornos reales.

Apache Spark: (Meng et al., 2015) es un *framework* de código abierto construido para el procesamiento rápido de Big Data. De fácil uso y análisis sofisticado. Spark tiene varias ventajas comparado con otros *frameworks* de procesamiento de Big Data que se basan en MapReduce como son Hadoop y Storm. Este *framework* ofrece de manera unificada funcionalidades para la gestión de los requerimientos de Big Data, incluyendo además conjuntos de datos de diversos tanto en naturaleza como en la fuente de los datos. Spark supera la velocidad de procesamiento de otros *frameworks* basados en MapReduce con capacidades como: almacenamiento de datos en memoria y procesamiento en tiempo real. Mantiene los resultados intermedios en memoria en lugar de escribirlos en disco, resulta muy útil especialmente cuando se necesita trabajar sobre el mismo conjunto de datos varias veces. Está diseñado para ser un motor de ejecución que puede trabajar en memoria o en disco.

## Propuesta

Se propone una plataforma de procesamiento distribuido de flujos de datos demográficos e históricos de compras procedentes de varias fuentes, basada en la interoperabilidad de las tecnologías antes descritas. Para ello se desplegó un clúster de procesamiento sobre un conjunto de computadoras de escritorio de gama media, en el cual se ejecuta SAMOA configurado con varios motores de procesamiento (SPE) entre ellos Storm y S4. Estos se encargan de tareas como la serialización de los datos, los cuales son evaluados en *Scalable Advanced Massive Online Analysis* (SAMOA) (Bifet y De Francisci Morales, 2014) utilizan los algoritmos de *Massive Online Analysis* (MOA) (Bifet, Holmes, Kirkby & Pfahringer, 2010) y WEKA (Witten & Frank, 2005) para las etapas de preprocesamiento y de modelado de los datos.

Fueron evaluados varios de los algoritmos disponibles en los *frameworks* antes mencionados, sobre el conjunto de datos *Bank Marketing Data Set* logrando identificar los grupos objetivos y proponiendo un modelo de respuesta predictivo en tiempo real.

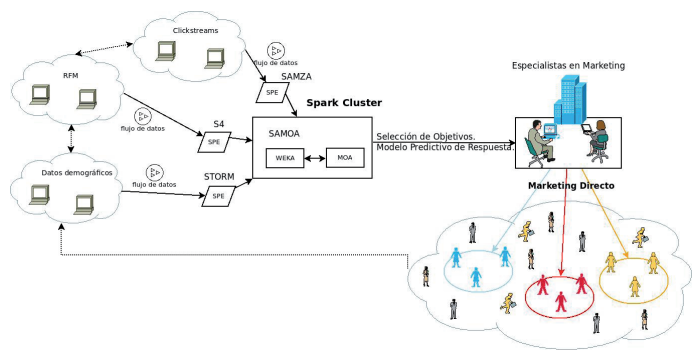

Figura 3. Plataforma propuesta.

## CONCLUSIONES

Con el desarrollo de las Tecnologías de la Informática y las Telecomunicaciones, las empresas están adoptando estrategias de marketing directo que permiten identificar a los clientes potenciales de sus productos y/o servicios, se evita así grandes gastos de recursos y esfuerzos en campañas publicitarias masivas. En este artículo fue descrita la naturaleza de los datos empleados por los algoritmos para la selección de grupos objetivos de clientes, al destacar conjuntos de datos actualizados. Además se propuso una plataforma para el procesamiento de datos de clientes provenientes de diferentes fuentes. Para ello se evaluaron tecnologías para ambientes de flujos de datos distribuidos, se posibilita así el análisis de los mismos en tiempo real. Varios criterios de selección se aplicaron, lo cual garantiza la interoperabilidad, y las libertades que ofrecen los productos de *Software* Libre. Sobre esta propuesta pueden desarrollarse y evaluarse nuevos algoritmos que requieren de un procesamiento de flujos de datos. De esta manera se facilitan tareas como la simulación de los entornos distribuidos, la serialización de los datos, entre otras que constituyen requerimientos funcionales para investigaciones relacionadas con este campo de acción.

## REFERENCIAS BIBLIOGRÁFICAS